\documentstyle[12pt]{article}

\def\abstract#1{\vskip 7mm 
        \begin{center}{\large Abstract}\par \smallskip
                \begin{minipage}[c]{12cm}
                        \small #1
                \end{minipage}
        \end{center}
}
\def\title#1{\begin{center}{\Large\bf #1}\end{center}}
\def\author#1{\vskip 5mm \begin{center}{#1}\end{center}}
\def\address#1{\begin{center}{\it #1}\end{center}}
\newcommand{\beq}{\begin{equation}}
\newcommand{\eeq}{\end{equation}}
\newcommand{\beqa}{\begin{eqnarray}}
\newcommand{\eeqa}{\end{eqnarray}}
\makeatletter
\@ifundefined{lesssim}{}{}
\@ifundefined{gtrsim}{}{}
\def\vereq#1#2{\lower3pt\vbox{\baselineskip1.5pt \lineskip1.5pt
\ialign{$\m@th#1\hfill##\hfil$\crcr#2\crcr\sim\crcr}}}
\makeatother

\begin{document}

\title{%
Constancy of the Constants of Nature
\footnote{based on a talk presented at {\it Frontier of Cosmology and 
Gravitation}, YITP, Kyoto, April 25-27, 2001.}
}
\author{%
  Takeshi Chiba\footnote{E-mail:chiba@tap.scphys.kyoto-u.ac.jp}
}
\address{%
  Department of Physics, Kyoto University,\\
   Kyoto, 606--8502, Japan
}

\abstract{
The current observational and experimental bounds on  time 
variation of the constants of Nature are briefly reviewed.
}

\section{Introduction}

\subsection{History}

To our knowledge, Dirac appears to have been the first who argued 
for the possibility of  time variation of the constants of nature. 
As is well-known, dimensionless 
numbers involving $G$ are huge (or minuscule). For example, the ratio 
of the electrostatic force to the gravitational force  between an electron 
and a proton is \footnote{We use the units of $\hbar =c=1$.}
\beq
N_1={e^2\over Gm_pm_e}\simeq 2\times 10^{39},
\eeq
where $e$ is the electric charge, $m_p$ is the proton mass and $m_e$ is 
the electron mass. 
Similarly, the ratio of the Hubble horizon radius of the Universe, 
$H_0^{-1}$ to the classical radius of an electron is
\beq
N_2={H_0^{-1}\over e^2m_e^{-1}}\simeq 3\times 10^{40}h^{-1},
\eeq
where $h$ is the Hubble parameter in units of ${\rm km s^{-1}Mpc^{-1}}$.
Curiously, the two nearly coincides, which motivated Dirac to postulate 
the so-called the large number hypothesis \cite{dirac38}. 
In his article entitled ``A new basis for cosmology'', he describes 
\cite{dirac38} 
\begin{quote}
{\it Any two of the very large dimensionless numbers occurring in Nature are
 connected by a simple mathematical relation, in which the coefficients are 
of the order of magnitude unity.}
\end{quote}
Thus if the (almost) equality $N_1={\cal O}(1)\times N_2$ holds 
always, then  $G$ must decrease with time $G\propto t^{-1}$\cite{dirac37}, 
or the fine structure constant, $\alpha\equiv e^2$, must increase with time 
$\alpha \propto t^{1/2}$ \cite{gamov67} since $H \propto t^{-1}$.

Nowadays we know that such a huge dimensionless number like $N_1$ is 
related to the gauge hierarchy problem. In fact, the gauge couplings 
 are {\it running} (however, only logarithmically) 
as the energy grows, and all the gauge couplings are believed to unify at 
the fundamental energy scale (probably string scale). 
The fact that $N_1$ nearly coincides with $N_2$ may be just accidental, and 
pursuing the relation between them is numerological speculation (or requires 
anthropic arguments). However, 
Pandora's box was opened. 

\subsection{Modern Motivation}

String theory is the most promising approach to unify all the fundamental 
forces in nature. It is believed that in string theory all the 
coupling constants and parameters (except the string tension) in nature are 
derived quantities and are determined by the vacuum expectation values of 
the dilaton and moduli. 

On the other hand, we know that the Universe is expanding. Then it is 
no wonder to imagine the possibility of  time variation of the 
constants of nature during the evolution of the Universe. 

In fact, it is argued that the effective potentials of dilaton or moduli 
induced by nonperturbative effects may exhibit runaway structure; 
they asymptote zero for the weak coupling limit where dilaton becomes minus 
infinity or internal radius becomes infinity and symmetries are restored 
in the limit \cite{dine85,witten00}. Thus it is expected that as these 
fields vary, the natural ``constants'' may change in time 
and moreover the violation of the weak equivalence principle may be induced 
\cite{witten00,damour94} (see also \cite{marciano84,maeda88} 
for earlier discussion). 

In this article, we review the current experimental (laboratory, 
astrophysical and geophysical) constraints on  
time variation of the constants of nature. In particular, we consider 
$\alpha$, $G$, and $\Lambda$. We assume $H_0=100h {\rm km/s/Mpc}$ with 
$h=0.65$ for the Hubble parameter and $\Omega_M=0.3$ and 
$\Omega_{\Lambda}=0.7$ for the cosmological parameters. 
For earlier expositions, see \cite{dyson72,bt86} for example.

\section{$\alpha$}

In this section, we review the experimental constraints on time variation 
of the fine structure constant. The results are summarized in Table 1.

\subsection{Oklo Natural Reactor and $\dot\alpha$}

In 1972, the French CEO (Commissariat \`a l'Energie Atomique) discovered 
ancient natural nuclear reactors in the ore body of the Oklo uranium mine 
in Gabon, West Africa. 
It is called the Oklo phenomenon. The reactor operated about 1.8 billion 
years ago corresponding to $z\simeq 0.13$ for the assumed cosmology 
($h=0.65,\Omega_M=0.3,\Omega_{\Lambda}=0.7$).

Shlyakhter noticed the extremely low resonance energy ($E_r=97.3 {\rm meV}$) 
of the reaction
\beq
~^{149}S_m+n\rightarrow ~^{150}S_m+ \gamma, 
\eeq
and hence the abundance of $~^{149}S_m$ (one of the nuclear fission 
products of $~^{235}U$) observed at the Oklo can be a good 
probe of the variability of the coupling constants \cite{shlyakhter76}.
The isotope ratio of $~^{149}S_m/~^{147}S_m$ is $0.02$ rather 
than $0.9$ as in natural samarium due to the neutron flux onto $~^{149}S_m$ 
during the uranium fission. 
{}From an analysis of nuclear and geochemical data, the operating conditions 
of the reactor was inferred and the thermally averaged neutron-absorption 
cross section could be estimated. 
The nuclear Coulomb energy is of order $V_0 \sim 1 {\rm MeV}$, and 
its change $\Delta V_0$ is related to the change of $\alpha$ as
\beq
{\Delta V_0\over V_0} = {\Delta E_r\over V_0}= 
{\Delta \alpha \over \alpha}.
\eeq
By estimating the uncertainty in the resonance energy, Shlyakhter
obtained $\dot \alpha/\alpha=10^{-17}{\rm yr}^{-1}$. 
Damour and Dyson reanalysed the data and 
obtained $(-6.7\sim 5.0)\times 10^{-17}{\rm yr}^{-1}$ \cite{damour96}. 
Using new samples that were carefully collected to minimize natural 
contamination and also on a careful temperature estimate of the reactors, 
Fujii et al. reached a tighter bound\footnote{They noted that data is 
also consistent with a non-null result: 
$(-4.9\pm 0.4)\times 10^{-17}{\rm yr}^{-1}$, indicating an apparent 
evidence for the time variability. However, from the analysis of the 
isotope compositions of $G_d$, the consistency of the $S_m$ and $G_d$
results supports the null results.}
$(-0.2\pm 0.8)\times 10^{-17}{\rm yr}^{-1}$ 
\cite{fujii00}.

\subsection{(Hyper)Fine splitting and $\dot\alpha$}

Since fine structure levels depend multiplicitely on $\alpha$, 
wavelength spectra of cosmologically distant quasars provide a natural 
laboratory  for investigating the time variability of $\alpha$. 
Narrow lines in quasar spectra are produced by absorption of radiation in 
intervening clouds of gas, many of which are enriched with heavy elements. 
For example, the separation between the wavelength corresponding to the 
transition $^2S_{1/2} \rightarrow ^2P_{3/2}$ and the wavelength 
corresponding to the transition $^2S_{1/2} \rightarrow ^2P_{1/2}$ 
in an alkaline ion is proportional to $\alpha^2$. 
Because quasar spectra contain doublet 
absorption lines at a number of redshifts, it is possible to check for 
time variation in $\alpha$ simply by looking for changes in the doublet 
separation of alkaline-type ions with one outer electron as a function of 
redshift \cite{bahcall65,wolfe76}. 
By looking at Si IV doublet, Cowie and Songaila obtained the limit up to 
$z\simeq 3$: $|\Delta \alpha/\alpha| < 3.5\times 10^{-4}$ \cite{cowie95}.
Also by comparing the hyperfine 21 cm HI transition with optical atomic 
transitions in the same cloud at $z\simeq 1.8$, 
they obtained a bound on the fractional change in $\alpha$ up to 
redshift $z\simeq 1.8$: 
$\Delta \alpha/\alpha = (3.5\pm 5.5)\times 10^{-6}$, corresponding to 
$\dot\alpha/\alpha   =(-3.3\pm 5.2)\times 10^{-16}{\rm yr}^{-1}$ 
\cite{cowie95}. Recently, by comparing the absorption by the HI 21 cm 
hyperfine transition  (at $z=0.25,0.68$) with the absorption by 
molecular rotational transitions,  Carilli et al. obtained a bound: 
$|\Delta \alpha/\alpha| < 1.7\times 10^{-5}$ \cite{carilli00}. 

Webb et al. \cite{webb99} introduced a new technique (called many-multiplet 
method) that compares the absorption wavelengths of magnesium 
and iron atoms in the same absorbing cloud, which is far more 
sensitive than the alkaline-doublet method. 
They observed a number of intergalactic clouds at redshifts from 0.5 to 1.6. 
For the entire sample they find 
$\Delta \alpha/\alpha = (-1.1\pm 0.4)\times 10^{-5}$, consistent with a 
null result within 3 $\sigma$. They noted that the deviation is dominated 
by measurements at $z>1$, where 
$\Delta \alpha/\alpha = (-1.9\pm 0.5)\times 10^{-5}$. 

Recently, Webb et al. \cite{webb00} pesented further evidence for time 
variation of $\alpha$ by reanalysing the previous data and including new 
data of Keck/HIRES absorption systems. The results indicate a smaller 
$\alpha$ in the past and the optical sample shows a 4 $\sigma$ deviation 
for $0.5 < z < 3.5$: 
$\Delta \alpha/\alpha = (-0.72\pm 0.18)\times 10^{-5}$. 
They noted that the potentially 
significant systemtic effects only make the deviation siginificant. 
If the result is correct, it would have profound implications for our 
understanding of fundamental physics.
So the claim needs to be verified independently by other observations.

\subsection{Laboratory Tests: Clock Comparison}

Laboratory constraints are based on clock comparisons with ultrastable 
oscillatiors of different physical makeup such as the superconducting 
cavity oscillator vs the cesium hyperfine clock transition \cite{turneaure76} 
or the Mg fine structure transition vs the cesium hyperfine clock transition 
\cite{godone93}. Since a hyperfine splitting is a function of $Z\alpha$, 
such a clock comparison can be a probe of 
time variation of $\alpha$. The clock comparisons place a limit on 
present day variation of $\alpha$ and are repeatable and hence are 
complementary to the geophysical or cosmological constraints. 
Comparisons of rates between clocks based on hyperfine transitions 
in alkali atoms with different atomic number $Z$ (H-maser and 
${\rm Hg^{+}}$ clocks) over 140 days yield a bound:  
$|\dot\alpha/\alpha| \leq 3.7\times 10^{-14}{\rm yr}^{-1}$ 
\cite{prestage95}.

\subsection{Cosmology and $\dot\alpha$: Big-Bang Nucleosynthesis and 
Cosmic Microwave Background}

\paragraph*{Big-Bang Nucleosynthesis.}

The $~^4{\rm H_e}$ abundance is primarily determined by the 
neutron-to-proton ratio prior to nucleosynthesis which before the freeze-out 
of the weak interaction rates at a temperature $T_f\sim 1$MeV, is given 
approximately by the equilibrium condition:  
\beq
(n/p)\simeq \exp(-Q/T_f), 
\label{yp}
\eeq
where $Q=1.29$MeV is the mass difference between neutron and proton. 
Assuming that all neutrons are incorporated into $~^4{\rm H_e}$, 
$~^4{\rm H_e}$ abundance $Y_p$ is given by
$Y_p \simeq 2(n/p)/[ 1+(n/p)]$.
The freeze-out temperature is determined by the competition between the weak 
interaction rates and the expansion rate of the Universe. The dependence of 
$T_f$ on  the weak and gravitational couplings is given by
\beq
T_f \propto G_F^{-2/3}G^{1/6},
\label{freeze}
\eeq
where $G_F$ is the Fermi constant.

Changes in $Y_p$ are induced by changes in $T_f$ and $Q$. However, it 
is found that $Y_p$ is  most sensitive to changes in $Q$ \cite{kolb86}.
The $\alpha$ dependence of $Q$ can be written as 
\cite{gasser,dixit88,campbell95}
\beq
Q \simeq 1.29 -0.76\times \Delta\alpha/\alpha ~~{\rm MeV}.
\eeq
Comparing with the observed $Y_p$, a bound on $\Delta\alpha/\alpha$ is 
obtained: $|\Delta\alpha/\alpha|\leq 2.6\times 10^{-2}$ \cite{dixit88}. 
Recently, it is argued that the presently unclear observational situation 
concerning the primordial abundances precludes a better bound than 
$|\Delta\alpha/\alpha|\leq 2\times 10^{-2}$ \cite{bergstrom99}.

\paragraph*{Cosmic Microwave Background.}

Changing $\alpha$ alters the ionization history of the universe and hence 
affects the spectrum of cosmic microwave background fluctuations: it changes 
the Thomson scattering cross section, $\sigma_T=8\pi \alpha^2/3m_e^2$, and 
hence the differential optical depth $\dot\tau$ of photons due to Thomson 
scattering through $\dot\tau =x_en_p\sigma_T$ where $x_e$ is the ionization 
fraction and $n_p$ is the number density of electrons; it 
also changes the recombination of hydrogen.

The last scattering surface is defined by the peak of  
the visibility function, $g(z)=e^{-\tau(z)}dt/dz$, which measures 
the differential probability 
that a photon last scattered at redshift $z$. 
As explained in \cite{hannestad99}, increasing $\alpha$ affects the 
visibility function $g(z)$: it increases the redshift of the last scattering 
surface and decreases the thickness of the last scattering surface. 
This is because the increase in $\alpha$ shifts $g(z)$ to higher redshift 
since the equilibrium ionization fraction, $x_e^{EQ}$, is shifted to 
higer redshift and because $x_e$ more closely tracks $x_e^{EQ}$ for larger 
$\alpha$. 

An increase in $\alpha$ changes the spectrum of CMB fluctuations: the peak 
positions in the spectrum shift to higher values of $\ell$ (that is, 
a smaller angle) and the values of $C_{\ell}$ increase \cite{hannestad99}. 
The former effect is due to  the increase of the redshift of the last 
scattering surface, while the latter is due to a larger early integrated 
Sachs-Wolfe effect because of an earlier recombination. It is concluded that 
the future of cosmic microwave background experiment (MAP, Planck)  could 
be sensitive to $|\Delta\alpha/\alpha|\sim 10^{-2}-10^{-3}$ 
\cite{hannestad99,avelino01}.\footnote{After this talk, new 
analyses of BOOMERanG \cite{netterfield} and MAXIMA \cite{lee} and the 
first year results from DASI \cite{pryke} appeared. In those data  
the second (and even the third) peak is now clearly seen, which 
discourages the motivation for non-standard recombination scenarios that 
were attracted attention regarding the interpretation of the apparent 
absence of the second peak in the previous data.}

\begin{table}
  \begin{center}
  \setlength{\tabcolsep}{3pt}
  \begin{tabular}{|l|c|c|r|} \hline
  &  redshift & $\Delta\alpha/\alpha$ & $\dot\alpha/\alpha({\rm yr}^{-1})$ \\ \hline
  Atomic Clock\cite{prestage95} & 0 & & $\leq 3.7\times 10^{-14}$ \\  \hline
   Oklo(Damour-Dyson\cite{damour96}) &  0.13 & $(-0.9\sim 1.2)\times 10^{-7}$ & $(-6.7\sim 5.0)\times 10^{-17}$   \\  \hline
   Oklo(Fujii et al.\cite{fujii00})  & 0.13  & $(-0.36\sim 1.44)\times 10^{-8}$ & $(-0.2\pm 0.8)\times 10^{-17}$\\  \hline
  HI 21 cm\cite{cowie95} & 1.8 & $(3.5\pm 5.5)\times 10^{-6}$ & $(-3.3\pm 5.2)\times 10^{-16}$ \\  \hline
  HI 21 cm\cite{carilli00} & 0.25,0.68 & $<1.7 \times10^{-5}$ &  \\  \hline
  QSO absorption line\cite{webb99} & $0.5-1.6$ & $(-1.1\pm 0.4)\times 10^{-5}$ &  \\  \hline
  QSO absorption line\cite{webb00} & $0.5-3.5$ & $(-0.72\pm 0.18)\times 10^{-5}$ &  \\  \hline
  CMB\cite{hannestad99} & $10^{3}$ &  $<10^{-2}\sim 10^{-3}$ & 
$< 10^{-12}\sim 10^{-13}$ \\  \hline
  BBN\cite{bergstrom99} & $10^{10}$ &  $< 2\times 10^{-2}$ & $< 1.4\times 10^{-12}$ \\  \hline
  \end{tabular}
  \end{center}
\caption{
Summary of the experimental bounds on  time variation of the fine 
structure constant. $\Delta\alpha/\alpha\equiv 
(\alpha_{\rm then}-\alpha_{\rm now})/\alpha_{\rm now}$.
A bound on $\Delta\alpha/\alpha$ for CMB is a possible bound.
}
\end{table}

\section{$G$}

In this section, we review the experimental constraints on time variation 
of the Newton constant. For more detailed review see \cite{gillies97}. 
The results are summarized in Table 2.

\subsection{Viking Radar-Ranging to Mars,  Lunar-Laser-Ranging and $\dot G$}

If we write the effective gravitational constant 
$G$ as $G=G_0+\dot G_0(t-t_0)$,  the effect of changing $G$ is readily seen 
through the change in the equation of motion:
\beq
{d^2{\bf x}\over dt^2}=-{GM{\bf x}\over r^3}=-{G_0M{\bf x}\over r^3}-
{\dot G_0\over G_0}{G_0M\over r}{{\bf x}(t-t_0)\over r^2}.
\eeq
Thus time variation of $G$ induces an acceleration term in addition to the 
usual Newtonian and relativistic ones, which would affect the motion of 
bodies, such as planets and binary pulsar. 

A relative distance between the Earth and Mars was accurately measured 
by taking thousands of range measurements between tracking stations 
of the Deep Space Network and Viking landers on Mars.
{}From a least-squares fit of the parameters of the solar system model to 
the data taken from various range measurements including those by 
Viking landers to Mars (from July 1976 to July 1982), a bound on $\dot G$ 
is obtained: $\dot G/G =(2\pm 4)\times 10^{-12}{\rm yr}^{-1}$ 
\cite{hellings83}.

Similarly, Lunar-Laser-Ranging measurements have been used to accurately 
determine parameters of the solar system, in particular the Earth-Moon 
separation. From the analysis of the data from 1969 to 1990, a bound is 
obtained: $\dot G/G =(0.1\pm 10.4)\times 10^{-12}{\rm yr}^{-1}$ 
\cite{muller91}; while from the data from 1970 to 1994, 
$\dot G/G =(1\pm 8)\times 10^{-12}{\rm yr}^{-1}$ \cite{williams96}.

\begin{table}
  \begin{center}
  \setlength{\tabcolsep}{3pt}
  \begin{tabular}{|l|c|c|r|} \hline
  &  redshift & $\Delta G/G$ & $\dot G/G({\rm yr}^{-1})$ \\ \hline
  Viking Lander Ranging\cite{hellings83} & 0 & & $(2\pm 4)\times 10^{-12}$ \\  \hline
  Lunar Laser Ranging\cite{williams96} & 0 & & $(1\pm 8)\times 10^{-12}$ \\  \hline
  Double Neutron Star Binary\cite{taylor91} &  0 & & $(1.10\pm 1.07)\times 10^{-11}$   \\  \hline
  Pulsar-White Dwarf Binary\cite{taylor94} &  0 & & $(-9\pm 18)\times 10^{-12}$   \\  \hline
   Helioseismology\cite{krauss98}  & 0  & & $<1.6\times 10^{-12}$ \\  \hline
  Neutron Star Mass\cite{thorsett96} &  $0-3\sim 4$ & & $(-0.6\pm 2.0)\times 10^{-12}$   \\  \hline
  BBN\cite{krauss90} & $10^{10}$ &  $ -0.3\sim 0.4$ & $(-2.7\sim 2.1)\times 10^{-11}$ \\  \hline
  \end{tabular}
  \end{center}
\caption{
Summary of the experimental bounds on time variation of the gravitational 
constant. $\Delta G/G\equiv 
(G_{\rm then}-G_{\rm now})/G_{\rm now}$.
}
\end{table}

\subsection{Binary Pulsar and $\dot G$}

The timing of the orbital dynamics of binary pulsars provides 
a new test of  time variation of $G$. 
To the Newtonian order, the orbital period of a two-body system is given by
\beq
P_b=2\pi \left({a^3\over Gm}\right)^{1/2}=
{2\pi \ell^3\over G^2m^2(1-e^2)^{3/2}},
\eeq
where $a$ is the semi-major axis, $\ell=r^2\dot\phi$ is the angular 
momentum per unit mass, $m$ is a Newtonian-order mass parameter, and 
$e$ is the orbital eccentricity. This yields the orbital-period 
evolution rate 
\beq
{\dot P_b\over P_b}=-2{\dot G\over G}+3{\dot\ell\over \ell}-2{\dot m\over m}.
\eeq
Damour, Gibbons and Taylor showed that the appropriate phenomenological 
limit on $\dot G$ is obtained by
\beq
{\dot G\over G}=-{\delta\dot P_b\over 2P_b},
\eeq
where $\delta\dot P_b$ represents whatever part of the observed orbital 
period derivative that is not otherwise explained \cite{damour88}. 
{}From the timing of the binary pulsar PSR 1913+16, a bound is obtained: 
$\dot G/G= (1.0\pm 2.3)\times 10^{-11}{\rm yr}^{-1}$ \cite{damour88} 
(see also \cite{taylor91} and \cite{taylor94}). 
However, only for the orbits of bodies which have negligible gravitational 
self-energies, the simplifications can be made that $\dot P_b/ P_b$ is 
dominated by $-2\dot G/G$ term. When the effect of the variation in 
the gravitational binding energy induced by a change in $G$ is taken into 
account, the above bound is somewhat weakened depending on the equation 
of state \cite{nordtvedt90}.

\subsection{Stars and $\dot G$}

Since gravity plays an important role in the structure and evolution of 
a star, a star can be a good probe of time variation of $G$.  
It can be shown from a simple dimensional analysis that the luminosity 
of a star is proportional to $G^7$ \cite{teller48}. Increasing  
$G$ is effectively the same, via the Poisson equation, as increasing 
the mass or average density of a star, which increases its average mean 
molecular weight and thus increases the luminosity of a star. 
Since a more luminous star burns more hydrogen, 
the depth of convection zone is affected which is determined directly 
from observations of solar $p$-mode (acoustic wave) spectra \cite{convec91}.
Helioseismology enables us probe the structure of the solar interior. 
Comparing the $p$-mode oscillation spectra of varying-$G$ solar models with 
the solar $p$-mode frequency observations, a tighter bound on $\dot G$ is 
obtained: $|\dot G/G|\leq 1.6\times 10^{-12}{\rm yr}^{-1}$ \cite{krauss98}.

The balance between the Fermi degeneracy pressure of a cold electron gas 
and the gravitational force determines the famous Chandrasekhar mass
\beq
M_{Ch} \simeq G^{-3/2}m_p^{-2},
\eeq
where $m_p$ is the proton mass. 
Since $M_{Ch}$ sets the mass scale for the late evolutionary stage of 
massive stars, including the formation of neutron stars in core collapse 
of supernovae, it is expected that the average neutron mass is given by 
the Chandrasekhar mass. Measurements of neutron star masses and ages over 
$z <3\sim 4$ yield a bound, 
$\dot G/G=(-0.6\pm 2.0)\times 10^{-12}{\rm yr}^{-1}$ \cite{thorsett96}.

\subsection{Cosmology and $\dot G$: Big-Bang Nucleosynthesis}

The effect of changing $G$ on the primordial light abundances (especially 
$~^4{\rm H_e}$) is already seen in Eq.(\ref{yp}) and Eq.(\ref{freeze}): 
an increase in $G$ increases the expansion rate of the universe, which 
shifts the freeze-out to an earlier epoch and results in a higher abundance 
of $~^4{\rm H_e}$. In terms of the ``speed-up factor'', $\xi\equiv H/H_{SBBN}$,
$Y_p$ is well fitted by \cite{walker91}
\beq
Y_p\simeq 0.244 + 0.074 (\xi^2-1).
\eeq
If $Y_p$ was between $0.22$ and $0.25$, then $-0.32< \Delta G/G < 0.08$, which 
corresponds to $\dot G/G=(-0.55\sim 2.2)\times 10^{-11}{\rm yr}^{-1}$. 
A similar (more conservative) bound was obtained 
in \cite{krauss90}: $-0.3< \Delta G/G < 0.4$.

\subsection{Recent Developments on $G_0$}

No laboratory measurements of $\dot G/G$ has been performed recently (see 
\cite{gillies97} for older laboratory experiments). This is mainly 
because the measurements of the present gravitational constant $G_0$ itself 
suffer from systematic uncertainties and have not been performed with good 
precision. 

Recently, Gundlach and Merkowitz measured $G_0$ with a torsion-balance 
experiment in which string-twisting bias was carefully eliminated 
\cite{gundlach00}. The result was a value of 
$G_0=(6.674215\pm 0.000092)\times 10^{-11}{\rm m^3kg^{-1}s^{-2}}$.
\footnote{Only recently, the measurement of $G$ 
with a torsion-strip balance resulted in 
$G_0=(6.67559\pm 0.00027)\times 10^{-11}{\rm m^3kg^{-1}s^{-2}}$,
 which is 2 parts in $10^4$ higher than the 
result of Gundlach and Merkowitz \cite{quinn01}. Probably the difference is 
still due to systematic errors hidden in one or both of the measurements.}
As the accuracy of the measurements improves, it may be possible to 
 place a bound on a present-day variation of $G$. It is important to 
pursue laboratory measurements of $\dot G/G$ since they 
are repeatable and hence are complementary to astrophysical and geophysical 
constraints. 

\section{$\Lambda$ or Dark Energy}

Finally, we briefly comment on the potential variability of the 
cosmological constant (or dark energy) because in the runaway scenario of 
dilaton or moduli $\phi$, $\dot\alpha/\alpha$ and 
$\dot G/G$ would close to $\dot\phi/\phi$ \cite{witten00}. 

\subsection{Evidence for $\Lambda >0$}

There are two arguments for the presence of dark energy. The first indirect 
evidence comes from the sum rule in cosmology:
\beq
\sum \Omega_i=1,
\eeq
where $\Omega_i\equiv 8\pi G\rho_i/3H^2_0$ is the density parameter of the 
i-th energy component, $\rho_i$. The density parameter of the curvature, 
$\Omega_K$, is defined by  $\Omega_K\equiv -k/a^2H^2_0$. Since the current 
obervational data indicate that matter density is much less than the 
critical density $\Omega_M<1$ and that the Universe is flat, we are led to 
conclude that the Universe is dominated by dark energy, 
$\Omega_{DE}=1-\Omega_M-\Omega_K>0$.

The second evidence for dark energy is from the observational evidence for the
accelerating universe \cite{sn}:
\beq
{\ddot a\over aH_0^2}=-{1\over 2}\left(\Omega_M(1+z)^3+
(1+3w)\Omega_{DE}(1+z)^{3(1+w)}\right)>0,
\eeq
where $w$ is the equation of state of dark energy, $w\equiv p_{DE}/\rho_{DE}$.
Since distance measurements to SNIa strongly indicate the Universe is 
currently accelerating, the Universe should be dominated by dark energy with 
negative pressure ($w<0$). We note that another argument for negative 
pressure comes from the necessity of the epoch of the matter domination.

\subsection{Supernova and $\dot\Lambda$}

A current bound on the equation of state of dark energy 
from supernova data is $w\leq -0.6$ \cite{sneos}.
Future observations of high redshift supernovae/galaxies/clusters 
would pin down the bound on $w$ to $w\leq -0.9$.
The extent of  time variation of dark energy density is readily seen from 
the equation of motion:
\beq
{\dot\rho_{DE}\over\rho_{DE}}=-3(1+w)H.
\eeq

\section{Conclusion}

A short account of the experimental constraints on the time variability of 
the constants of nature ($\alpha$ and $G$)) was given. 
Since there are some theoretical motivations for the time variability of 
the constants of nature and the implications of it are profound, 
refining these bounds remains important, and continuing searches for 
the possible time variability of the constants of nature should be made.

~

{\bf Acknowledgement} 
This work was supported in part by a Grant-in-Aid for Scientific 
Research (No.13740154) from the Japan Society for the Promotion of Science.

\end{document}